# Scale-dependent non-affine elasticity of semiflexible polymer networks

M. Atakhorrami[1,4], G.H. Koenderink[2,4], J.F. Palierne[3], F.C. MacKintosh[4], and C.F. Schmidt[4,5,*]

[1] present address : Home and Personal Healthcare, Philips Group Innovation - Europe, Cambridge, CB4 0FY, United Kingdom
[2] FOM Institute AMOLF, 1098 XG Amsterdam, The Netherlands
[3] École Normale Supérieure de Lyon, Laboratoire de Physique, Lyon, France
[4] Department of Physics and Astronomy, VU University, 1081 HV Amsterdam, The Netherlands
[5] Third Institute of Physics, Georg August University, 37077 Göttingen, Germany

*Author to whom correspondence should be addressed. E-mail: cfs@physik3.gwdg.de

## Abstract

The cytoskeleton of eukaryotic cells provides mechanical support and governs intracellular transport. These functions rely on the complex mechanical properties of networks of semiflexible protein filaments. Recent theoretical interest has focused on mesoscopic properties of such networks and especially on the effect of local, non-affine bending deformations on mechanics. Here, we study the impact of local network deformations on the scale-dependent mobility of probe particles in entangled networks of semiflexible actin filaments by high-bandwidth microrheology. We find that micron-sized particles in these networks experience two opposing non-continuum elastic effects: entropic depletion reduces the effective network rigidity, while local non-affine deformations of the network substantially enhance the rigidity at low frequencies. We show that a simple model of lateral bending of filaments embedded in a viscoelastic background leads to a scaling regime for the apparent elastic modulus $G'(\omega) \sim \omega^{9/16}$, closely matching the experiments. These results provide quantitative evidence for how different a semiflexible polymer network can feel for small objects, and they demonstrate how non-affine bending deformations can be dominant for the mobility of vesicles and organelles in the cell.

A major challenge for a quantitative understanding of the dynamics of living cells is their structural and dynamic complexity. Cells are highly structured composites containing a multi-component protein polymer network, the cytoskeleton, which is anchored to the cell membrane. The cytoskeleton is built of semiflexible protein polymers, mainly filamentous (F-) actin, microtubules and intermediate filaments, with large persistence lengths, about 9-15 μm in the case of F-actin [1-3]. The characteristic length scales of the cytoskeleton span many orders of magnitude, from the monomer size of tubulin or actin of ~50 Å via the network mesh size of ~10-100 nm up to the filament persistence length of microns to millimeters. The mechanical response of the cytoskeleton therefore depends on both the time and the length scale on which it is probed. Large-scale deformations of whole cells such as in morphogenetic events in developing embryos are determined by macroscopic viscoelasticity. By contrast, vesicles and organelles, which are transported along and through the cytoskeleton by motor proteins, are typically only tens of nanometers to micrometers in size, and therefore experience local response. Local elastic moduli inside cells have been measured by video microrheology, which analyzes fluctuations of embedded micron-sized probe particles [4]. However, it is difficult to interpret such measurements due to the structural and biochemical complexity of cells and to the intrinsic non-equilibrium fluctuations which confound these measurements [4, 5]. To circumvent these complexities, many microrheological studies have been performed on reconstituted cytoskeletal networks, in particular F-actin. In spite of the greater simplicity of these systems, we still do not fully understand the determinants of local viscoelastic responses. Studies have shown that local elastic properties of even simple F-actin networks differ from bulk elastic behavior when measured with particles that are small compared to the length of the actin filaments [6-11]. The physical origin of this discrepancy is not understood. The large persistence length of the cytoskeletal polymers implies that filament bending will be part of the local response. Such non-affine elastic deformations have been mainly treated theoretically and in simulations to date [12-17]. There has been some experimental evidence that non-affine deformations affect the response of F-actin networks to macroscopic shear [18], but the impact of nonaffinity on local mechanical properties is unknown.

Here we combine bulk rheology with optical microrheology measurements using optical trapping of micron-sized beads and laser interferometry to study the scale-dependent mechanics of cytoskeletal F-actin networks over a wide frequency range.We use a dual optical tweezers setup which enables us to measure in the same experiment two-particle microrheology (2pMR), which infers bulk properties from the correlated displacements of (distant) particle pairs [7, 19], and one-particle microrheology (1pMR), which is sensitive to local perturbations of the network by the probe particle itself.

We find evidence for two types of local non-continuum elastic effects. The probe particles introduce local pockets of reduced polymer concentration caused by entropic depletion, leading to a reduction of the local modulus compared to the bulk modulus at high frequencies. Surprisingly, however, the local elastic modulus at low frequencies, however, is enhanced. Such an effect has been suggested to exist on theoretical grounds [20] as a result of bending deformations of the semi-flexible filaments, but has never been experimentally demonstrated. We explain our observations by a theoretical model, which accounts for a transition from non-affine bending deformations of actin filaments on small scales to larger-scale affine stretching deformations of filaments. Non-affine filament bending leads to an intermediate scaling regime $G'(\omega) \sim \omega^{9/16}$, closely matching the experiments.

Samples were prepared by polymerizing rabbit skeletal muscle actin in a polymerization buffer containing 50 mM KCl, 2 mM $MgCl_2$, and 1 mM $Na_2ATP$ [21] (all chemicals from Sigma/Aldrich, USA). Actin networks were left to polymerize for at least 1 hour in glass/double-stick tape chambers in the presence of silica probe spheres of different diameters (Kisker Biotech, Germany, $D$ = 0.52 1.16, 2.56, and 5.0 µm, polydispersity 5% in all cases). The probe particles were larger than the average mesh size of the actin networks, $\xi \approx 0.3/\sqrt{c_A}$ (in µm, with concentration in mg/ml) [22]. The actin concentration $c_A$ ranged between 0.5 and 2 mg/ml. Bulk rheology measurements were performed in a custom-built piezorheometer [23, 24] by oscillating one glass plate sinusoidally in the direction normal to the surface with an amplitude of 1 nm and a frequency between 0.1 Hz and 10 kHz. Stress transmitted to a second parallel plate was measured by a second piezoelectric element. The squeezing flow leads to shear strain in the sample with an amplitude of ~$10^{-6}$, ensuring linear elastic response. Microrheology measurements were performed with a custom-built dual optical tweezers set-up [25]. Pairs of beads, at least 10 bead diameters away from any surface, were weakly trapped, using lasers with wavelengths of 1064 and 830 nm and trap stiffnesses below $3\times10^{-6}$ N/m. The thermal position fluctuations of both particles were detected by quadrant photodiodes at a sampling rate of 195 kHz and converted to storage and loss shear moduli using the fluctuation-dissipation theorem and a generalized Stokes relation [26]. 1pMR moduli represent averages over x and y displacements of single particles, while 2pMR moduli were derived from the correlated motions of two particles parallel to the line connecting their centers. Fluid inertia was negligible for the separation distances $r$ used here ($r = 10$ - $15\,\mu$m) [27, 28].

Storage and loss shear moduli $G'(\omega)$ and $G''(\omega)$ of actin solutions measured with 2pMR showed no significant dependence on probe-particle size. The 2pMR moduli agreed well with bulk moduli measured by piezorheometry (Figs. 1A, B). This confirms that 2pMR measures bulk properties when the separation distance between the two probe particles exceeds the relevant internal length scales of the network. In contrast, shear moduli obtained from individual particle displacements using 1pMR were significantly smaller than the bulk moduli (Figs. 1A, B). This was true even for the largest particles with a diameter of 5 µm, which are 17-fold bigger than the average mesh size (~300 nm) but still smaller than the average filament length (~20 µm).

To understand the origin of the difference between local and bulk elastic response, we compared results using probe particles of different sizes. When the particle diameter was reduced from 5.0 to 0.52 µm, the 1pMR loss modulus, $G''(\omega)$, decreased further in amplitude relative to the bulk modulus (Fig. 2A). This is in line with earlier 1pMR studies of actin networks [6-11]. At high frequencies, beyond 10 kHz, both the local and bulk loss moduli scaled with frequency according to $G''(\omega) \propto \omega^{3/4}$, as indicated in Fig. 1B. In this frequency regime, the viscoelastic response of the network is dominated by the relaxation of thermal bending fluctuations of individual filament segments between entanglement points [21, 29]. The moduli are therefore directly proportional to the concentration of polymer. The fact that local moduli were reduced compared to the bulk moduli suggests that the polymer concentration near the probe particles was locally reduced relative to the bulk density. The most likely explanation for this is entropic depletion: particles restrict conformations of polymers in their vicinity. Such an effect has also been observed for flexible synthetic polymers [30] and DNA [31]. The thickness of the depletion zone $\Delta$ can be estimated from a comparison of the local and bulk moduli at high frequencies [11, 31]. Our 1pMR data can be mapped onto the 2pMR data using a simplified model for depletion: each probe particle is assumed to be surrounded by a sharply defined, completely polymer-depleted shell of solvent of thickness $\Delta$ [19] (Fig. 1A, inset). Our data suggest that $\Delta$ grows with particle size (Fig. 3), consistent with previous reports [11]. Around spherical particles embedded in a solution of semiflexible polymers, $\Delta$ is expected to be of order the particle size [32, 33]. Near a flat wall $\Delta$ is predicted to be of order the polymer length or persistence length, whichever is shorter [34]. This depletion range is much larger than would be the case for solutions of entangled flexible polymers, where $\Delta$ is of order the mesh size [30, 31]. Local microrheology measurements for cytoskeletal polymer networks are therefore always affected by depletion, provided that the polymers are not attracted by the particles [35].

The local storage modulus $G'(\omega)$ measured with 1pMR exhibits a more complex dependence on probe particle size than the loss modulus. At high frequencies, $G'(\omega)$ measured with 1pMR is smaller than the bulk modulus, similar to $G''(\omega)$ (Fig. 3A). However, at low frequencies, the smallest particles, of 0.52 and 1.16 μm radii, experience a local elastic modulus exceeding the bulk modulus. Moreover, the local response exhibits an apparent elastic plateau not seen in the bulk data. This surprising observation suggests a second non-continuum effect arising from the interaction of the particles with the surrounding network. We hypothesize that this effect is due to non-affine local network deformations resulting from the transverse bending of filaments as the particle moves in the mesh. These transverse bends penetrate into the network over a range that is determined by a balance between their bending and the elastic deformation of the surrounding network [20]. As the particle moves, it collides with semiflexible filaments embedded in the rest of the network, considered here for simplicity as an elastic continuum. For a displacement of order $u$, the associated bending elastic energy is approximately $E_{bending} \sim n\kappa u^2/\lambda^3$, where $\lambda$ is the characteristic axial length over which the filament bends and $\kappa$ is its bending modulus. The number $n$ of bent filaments will increase with the particle size as $n \sim R/\xi$. In addition to the bending energy stored in the filament, there will also be energy stored in the distorted surrounding elastic continuum. If we assume that $\lambda \gg R$, this elastic energy is approximately $E_{elastic} \sim Gu^2\lambda$, where $G$ is the network shear modulus. Minimizing the total elastic energy $E = E_{elastic} + E_{bending}$ results in a characteristic deflection length [36],

$$\lambda_0 = \sqrt[4]{\frac{\kappa R}{G\xi}} \qquad (1)$$

A similar length scale $\lambda_0 = \sqrt[4]{\kappa/G}$ was found to be relevant for bending [37, 38] and buckling [39] of microtubules within the elastic cytoskeleton of cells. The probe particle in our model experiences an effective restoring force $f \sim -Gu\lambda_0$ due to the deformed filament. In increasingly stiff media, $\lambda_0$ will decrease and the assumption that it exceeds the particle radius $R$ will eventually break down. In that case the particle will experience the usual restoring force $f \sim -GuR$ of a particle in an elastic continuum. Thus, a transition is expected between a regime where the response is particle-size independent and dominated by non-affine filament bending to a regime where the length over which filaments are non-affinely bent is negligible compared to particle radius. The transition between these regimes occurs when $\lambda_0 \sim R$. Since the shear modulus of the embedding network increases with frequency, a transition of this kind should occur at a characteristic frequency $\omega_0$. To estimate this frequency, we assume that the network elastic modulus is $G = G^{(0)}$ in the plateau regime, which extends up to a frequency $\omega_e$, and has a power law form $G = Ac\omega^z$ with an amplitude $A$, above $\omega_e$. With these assumptions and using Eq. (1), the crossover frequency below which the filament bending effects should become apparent is of order

$$\omega_0 \sim \left(\frac{\kappa}{c\xi R^3}\right)^{1/z}. \qquad (2)$$

As long as $\lambda_0$ remains larger than $R$, the probe bead acts as if it were covered with rigid "bristles" due to its interactions with stiff filaments that tend to spread the force out. In other words, for frequencies $\omega < \omega_0$, the bead will behave as if it were of size $\lambda_0$. In this case, the 1pMR analysis, which assumes the actual particle radius $R$, will, in turn, report an apparently increased local modulus,

$$G_{local} \sim \frac{G\lambda}{R} \sim \frac{(c\omega)^{3z/4}}{R^{3/4}}.$$

(3)

Since for semiflexible actin networks $z = \frac{3}{4}$ [29], our model predicts that 1pMR reports a probe-size dependent local elastic modulus at low frequencies with a $\omega^{9/16}$ frequency dependence. This functional form should eventually cross over to the bulk response characterized by $\omega^{3/4}$ scaling.

Our experimental data are well explained by this model (Fig. 3B). The elastic modulus scales with frequency consistent with a $\omega^{9/16}$ law for the two smallest probes with diameters $D = 0.52$ and 1.16 µm over approximately 2 decades in frequency. For the intermediate-size particles ($D = 2.56$ µm) we find $\omega^{9/16}$ scaling over the whole frequency range. For the largest particles ($D = 5.0$ µm), the modulus scales as $G'(\omega) \sim \omega^{3/4}$ i.e. like the bulk modulus. The effect of non-affine deformations can be confirmed by the observation that all curves can be collapsed onto a master curve by simultaneously re-scaling $\omega$ by $\omega_0$ (Eq. (2)) and $G'(\omega)$ by the depletion effect. The scaled elastic moduli measured with different particle sizes collapse onto one curve, which scales as $\omega^{9/16}$ over a large frequency range and approaches $\omega^{3/4}$ at high frequencies (Fig. 3B). The scaling law can be emphasized (Fig. 3 inset) by plotting $\omega^{9/16}G'(\omega)$ against $\omega R^{12/3}$. The resulting curve is flat at low frequencies and crosses over to 3/16 scaling at high frequencies, consistent with $G' \sim \omega^{3/4}$.

Our results demonstrate that the local response of a semiflexible polymer network can be fundamentally different from the global response. By local probing with microscopic probes we could directly detect non-affine filament bending modes that are not measurable in macroscopic experiments. Local non-continuum effects are particularly relevant in biopolymer networks, where intrinsic length scales such as filament persistence length are on the micrometer scale. We have observed two prominent non-continuum effects, namely steric depletion of F-actin filaments in the vicinity of the probe particles and lateral bending deformations at low frequencies. We have introduced a simple model that accounts quantitatively for the frequency-dependent reduction in the mobility of the probes due to the local deformations of surrounding filaments. Interestingly, when translating the anomalous response to an apparent elastic modulus, this model predicts scaling as $\omega^{9/16}$, very close to the $\omega^{1/2}$ scaling that has frequently been reported [40, 41], but has thus far remained unexplained. An implication of our work is that it may not be appropriate to locally treat a semiflexible polymer network as a continuum, even when probe particles are substantially larger than the mesh size. This is crucial to understand before quantitative extensions of microrheology to the complex cytoskeleton of cells can be made. Furthermore it is important to realize that organelles or transport vesicles in the cytoplasm feel a response from the cytokeleton that is likely to be dominated by the local non-affine bending deformations of the rather stiff cytoskeletal polymers.

**Acknowledgments.**

We thank A.J. Levine, D.A. Head, F. Gittes, D.C. Morse, M. Pasquali, K.M. Addas, J. Liu, D.A. Weitz and P.A. Janmey for helpful discussions, J. van Mameren, J.I. Kwiecińska and M. Buchanan for software development, and K.C. Vermeulen for actin preparation. This work was supported by the Dutch Foundation for Fundamental Research on Matter (FOM). Additional support for C.F.S was provided by the German Science Foundation (DFG) Center for the Molecular Physiology of the Brain (CMPB) and by the DFG Collective Research Centers SFB 755 and SFB 937.

**Figure captions.**

**FIG. 1.** Local shear elastic moduli of entangled solutions of 1 mg/ml F-actin are smaller than bulk moduli: (A) storage moduli, $G'(\omega)$, (B) loss moduli $G''(\omega)$. Shear moduli obtained from 2pMR (black solid line: $D = 5.0$ μm; black dashed line: $D = 1.16$ μm ) agree with bulk rheology (open circles), while 1pMR measures smaller moduli (gray line: $D = 5.0$ μm).

**FIG. 2.** Loss moduli measured with 1pMR reveal local actin depletion near the probe beads (color online). (A) $G''(\omega)$ for a 1 mg/ml actin solution measured with 1pMR using particle sizes $D = 0.52$, 1.16, 2.56 and 5.0 μm (lines), and with bulk rheology (open circles). The line of slope 1 represents the solvent viscosity. Inset: schematic of two probe particles (radius $R$) at a distance $r$, surrounded by a depletion shell of thickness $\Delta$. (B) $\Delta$ calculated from microrheology data above 10 kHz plotted versus particle radius for different concentrations of actin, $c_A = 0.5$ (blue squares), 1 (red triangles) and 2 mg/ml (black circles). The dashed lines represent the mesh size for each concentration; the solid line is the particle radius $R$.

**FIG. 3.** Storage moduli, $G'(\omega)$, measured with 1pMR reveal local non-affine bending of actin filaments. (A) $G'(\omega)$ for a 1 mg/ml actin solution measured with 1pMR using particle sizes, $D = 0.52$, 1.16, 2.56 and 5.0 μm (lines), and with bulk rheology (open circles). The line of slope ¾ is the expected high frequency scaling, the line of slope 9 /16 is expected from local non-affine filament deformations. (B) Rescaled elastic modulus plotted versus frequency normalized by $2\pi R^{12/3}$, showing agreement with $\omega^{9/16}$ scaling (dashed line). Insert: same data with $G'(\omega)$ multiplied by $(\omega R^{12/3})^{-9/16}$, demonstrating $\omega^{9/16}$ scaling at low frequencies and $\omega^{3/4}$ scaling at high frequencies.

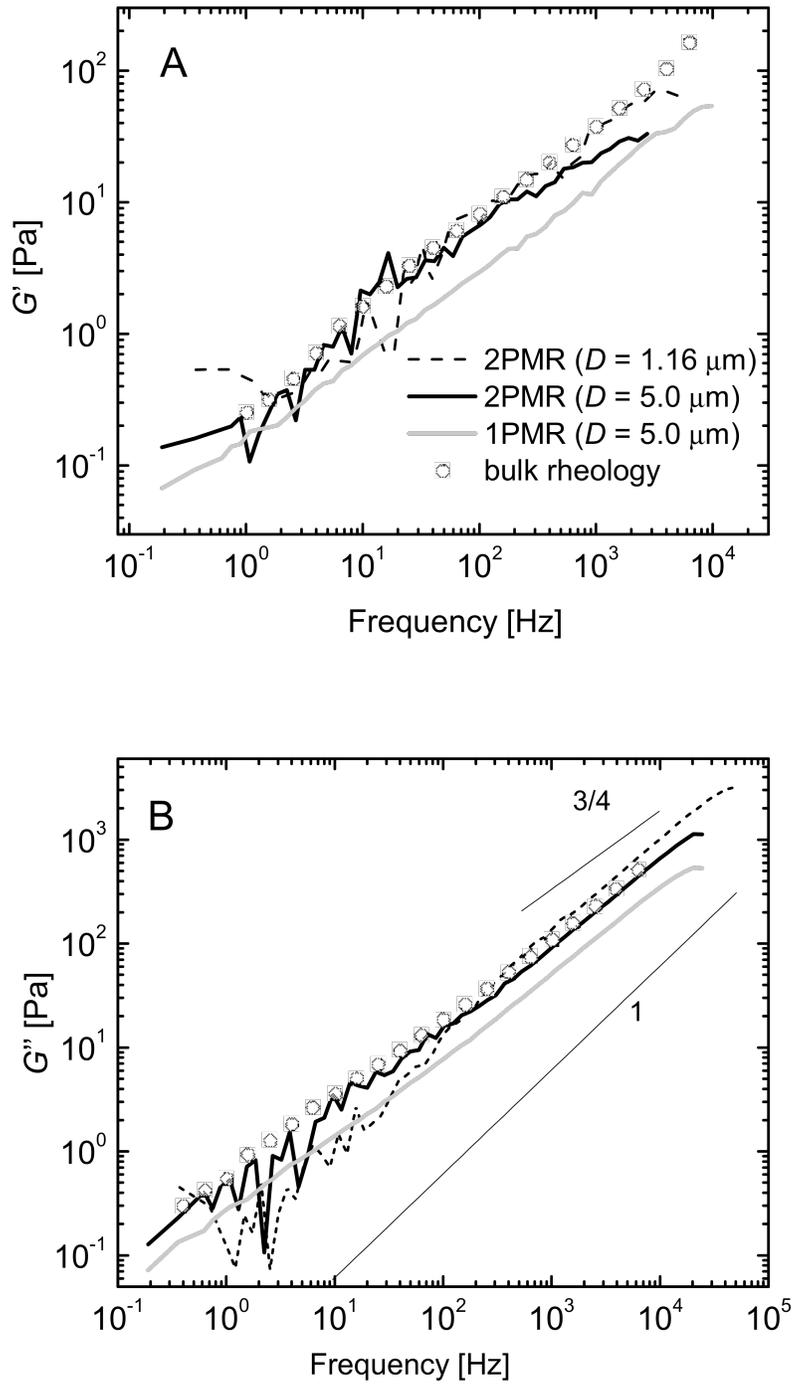

**FIG. 1 Atakhorrami *et al.*, 2012,** Scale-dependent non-affine elasticity of semiflexible polymer networks.

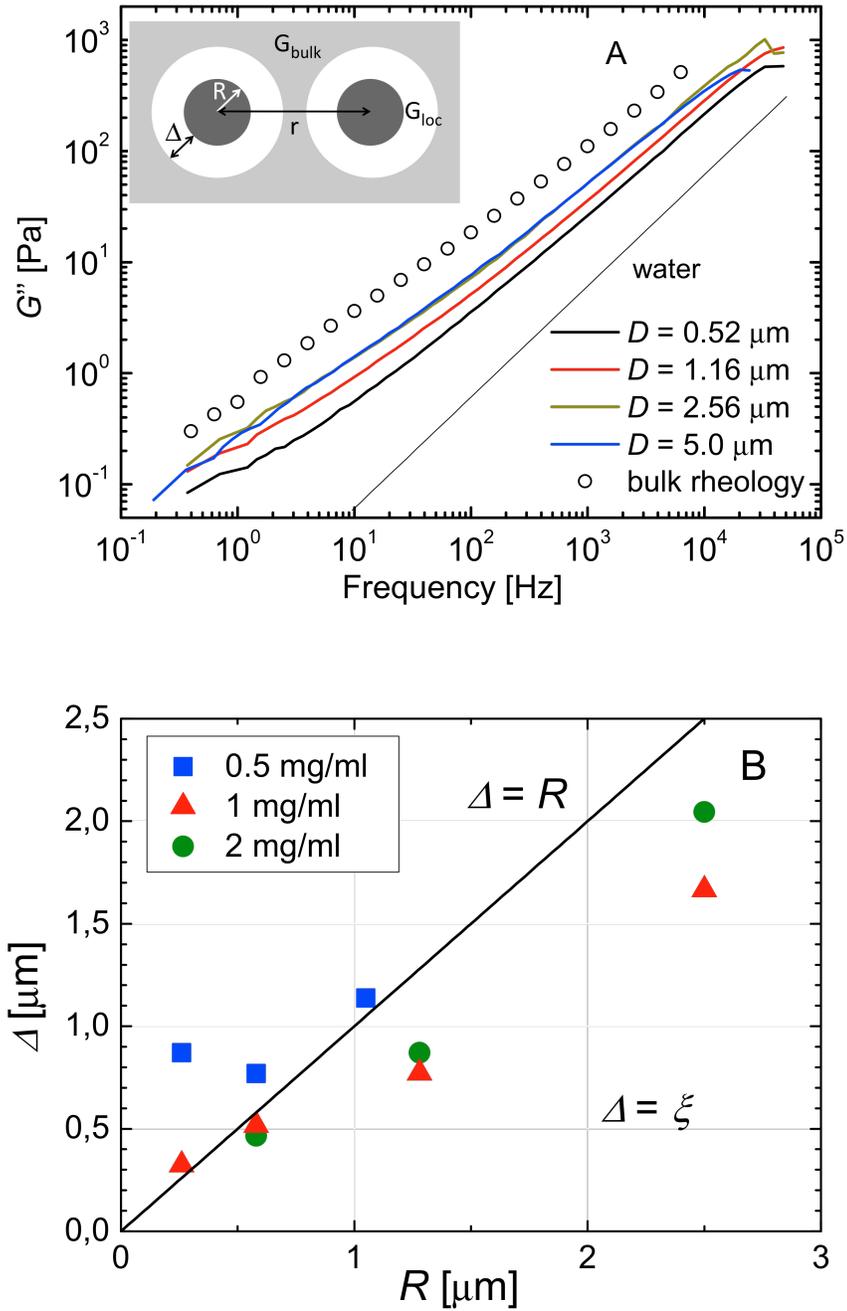

**FIG. 2**

**Atakhorrami *et al.*, 2012,** Scale-dependent non-affine elasticity of semiflexible polymer networks.

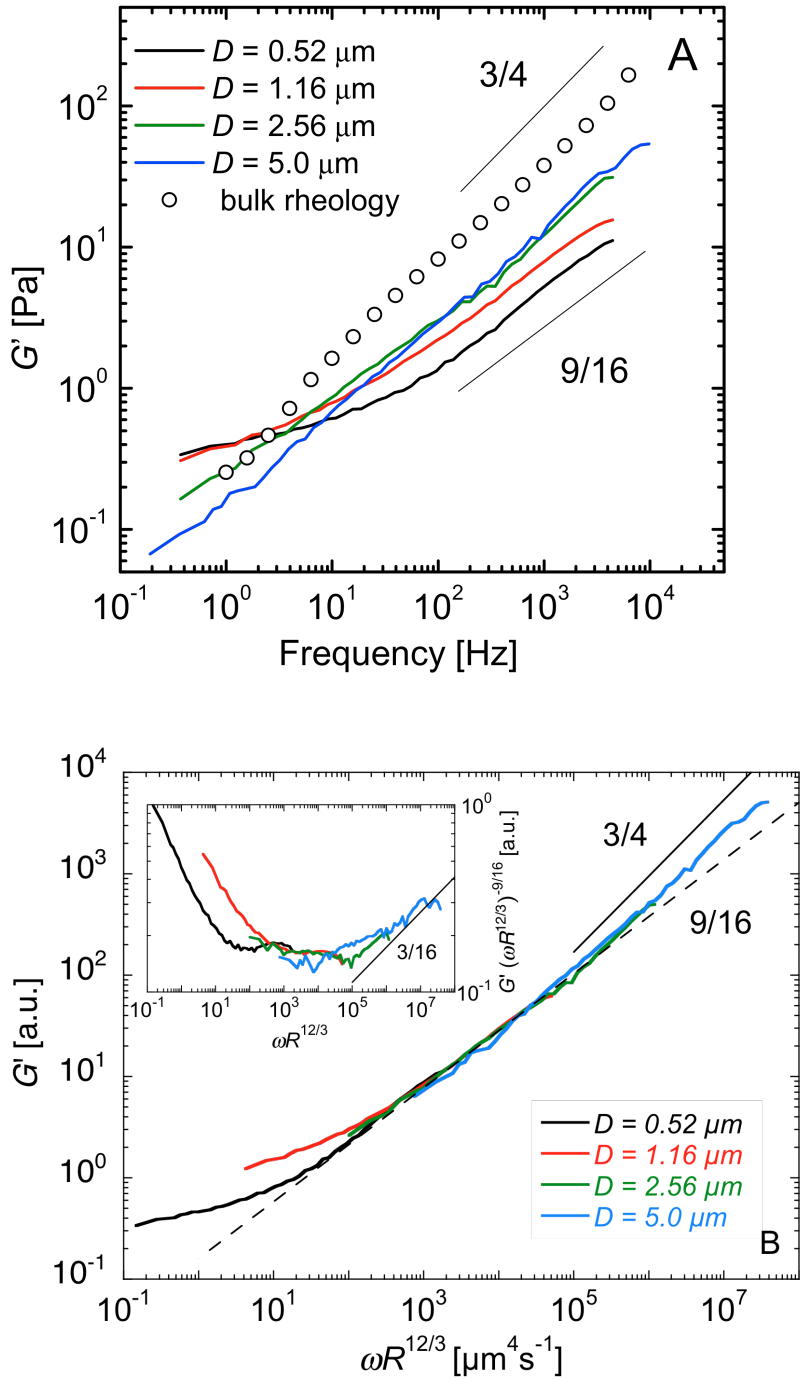

**FIG. 3 Atakhorrami *et al.*, 2012**, Scale-dependent non-affine elasticity of semiflexible polymer networks